\documentstyle[12pt]{article}

\newcommand{\ue}%
{\mbox{c\hspace{-0.4em}\rule[ 0.5ex]{0.3em}{0.04ex}\hspace{0.1em}}}
\newcommand{\Ue}%
{\mbox{C\hspace{-0.6em}\rule[ 0.75ex]{0.4em}{0.04ex}\hspace{0.2em}}}
\newlength{\signlength}%
\newcommand{\gs}%
{\settowidth{\signlength}{$<$}%
\raisebox{-0.35\signlength}%
{\makebox[1.7\signlength][c]{\makebox[0.pt]{$\sim$}%
\raisebox{0.6\signlength}{\makebox[0.pt]{$>$}}}}}
\newcommand{\ls}%
{\settowidth{\signlength}{$<$}%
\raisebox{-0.35\signlength}%
{\makebox[1.7\signlength][c]{\makebox[0.pt]{$\sim$}%
\raisebox{0.6\signlength}{\makebox[0.pt]{$<$}}}}}
\renewcommand{\ge}%
{\settowidth{\signlength}{$>$}%
\setlength{\unitlength}{0.1\signlength}%
\mbox{\begin{picture}(17,7.5)%
\linethickness{0.045\signlength}%
\put(0.0,1){\makebox(17,7.5){$>$}}%
\multiput(4.5,-1)(0.25,0.125){32}{\line(1,0){0.25}}%
\end{picture}}}
\renewcommand{\le}%
{\settowidth{\signlength}{$>$}%
\setlength{\unitlength}{0.1\signlength}%
\mbox{\begin{picture}(17,7.5)%
\linethickness{0.045\signlength}%
\put(0.0,1){\makebox(17,7.5){$<$}}%
\multiput(12.,-1)(-0.25,0.125){32}{\line(-1,0){0.25}}%
\end{picture}}}
\title{Scenario of superconducting transition for quasi-2D HTS}
\author{G.Sergeeva, NSC KIPT, Akademicheskaya 1, Kharkov 61108, Ukraine,}
\date{}
\begin{document}
\maketitle
\thispagestyle{empty}

\begin{abstract}

For
quasi-two-dimension HTS with spin-fluctuation pairing mechanism
the scenario of superconducting
transition is discussed.  The interaction of fluctuational spin waves with
holes in $Cu O_2$ planes in the mean field theory leads to the pairing of
holes and to the fluctuation generation of superconducting regions at
$T_{c0}$ (where $T_{c0}$ is the temperature of two dimension superconducting
transition) and also to essential temperature dependence of the strength of
interlayer coupling $t_{c}(T)$.  At sufficiently small values
of $t_{c}(T)$, the transition of the sample to the coherent
superconducting state occurs at $T_c< T_{c0}$.

\end{abstract}

  1. Layer compounds with weak interaction between the copper-oxygen planes
and with anisotropic resistivity ( for example, $\rho_c/ \rho_{ab} \sim
10^{5}$ for Bi2212 and $\rho_c/ \rho_{ab} \sim 10^{3}$ for $YBa_2Cu_3
O_{6.7}$ and $La_{2-x} Sr_x Cu O_4$) belong to quasi-2D
superconductors. The dependence of the resistivity $\rho_c(T)$  for such
 HTS's near the transition temperature $T>T_c$ has the semiconducting
character that is evidenced about noncoherent charge transfer along axis
$\widehat{c}$.
It's need to note that three dimension anisotropic
superconductors with coherent dynamics of charge in normal state becomes
quasi-2D ones under magnetic field $B>B_{cr}$ which is parallel to
axis $\widehat{c}$ (here $B_{cr}$ is the dimensional crossover field) or in
the case if the HTS is underdoped sample. In this paper following scenario
of superconducting transition is discussed for quasi-2D HTS with
spin-fluctuation pairing mechanism described by the generalized BCS theory 
[1-3]:

  1) in normal state the fluctuational spin excitations exist in $Cu O_2$
planes in the regions with measures which are confines by the correlation
length of the antiferromagnetic (AFM) fluctuations;

  2) the interaction of holes with fluctuational spin excitations leads
to the
fluctuational generation in $Cu O_2$ planes of superconducting regions
at the
temperature $T<T_{c0}$ where $T_{c0}$ is 2D superconducting transition
in the
mean field theory $T_{c0}$;

  3) the difference of the temperature dependencies of coherent lengths
$\xi _{ab} (T)$ in the $Cu O_2$ planes and out-of-plane
$\xi _c (T)$ leads to
anomal increase of $\rho_c(T)$ at the decreasing the temperature $T<T_{c0}$
and to essential temperature dependence of the
strength of interlayer coupling $t_{c}(T)$ [4-6];

  4) the transition to superconducting state with coherent dynamics of
 charge transfer occurs at enough small values of $t_c$, and the transition
temperature $T_c<T_{c0}$ is defined by the inequalities which at first were
received for layer systems in papers E.I.Kats [7] and L.N.Bulayevsky [8].

  2. Strong AFM fluctuations in quasi-two dimensions HTS are
prevented  to 2D Heisenberg ordering in copper-oxygen planes, in spite of
on essential anisotropy of exchange constants of in-plane and out-of-plane
interactions. In $Cu O_2$ planes long AFM order absences but
spin waves with linear dispersion are existing in the dielectric regions
with sizes which are confines by the correlation length of the
AFM fluctuations.
We can suppose that at $T \sim T_{c0}$ the exchange of
the holes by such spin excitations
leads to quasi-particles pairing with mechanism which is discussed nearly
10 years [1-3], and to the generation in $Cu O_2$ planes
of superconducting regions
with measures which are confined by the correlation length of the
AFM fluctuations. The temperature $T_{c0}$  defines by the mean value of
exchange interactions in $Cu O_2$ plane with taking to the account disrupted
couplings of copper spins in the dielectric regions.

  3. In temperature region, where  $\rho_c(T)$ are sensible over
 the Mott
limit, the transfer of charge along axis $\widehat{c}$  can consider as a
tunneling process of electron over nonconductor barrier $\rho_c \sim (N_0
t_{c})^{-1}$, where $N_0$ is the density of states in $Cu O_2$ planes.
The interest to the studying of the temperature
dependence $t_c$ is due by the inefficiency of
the attempts to account semiconductive
character of dependence $\rho_c(T)$ near $T_c$ by the decrease of state
density $N_0$ [9-10]. With decreasing of the temperature to $T< T_{c0}$ the
dependence $t_{c}(T)\simeq( \frac{\xi _c (T)}{\xi _{ab} (T)})^{2}$
is caused
by 2D-superconductive fluctuational effects, namely by the distinction of
temperature dependencies of coherent lengths $\xi _{ab}
(T)=\xi_{ab0}(1-T/T_{c0})^{-1/2}$in $Cu O_2$ plane and out-of-plane $\xi _c
(T)=\xi _{c0} (T)$ [6].

  At $T=T_c$ in quasi-2D HTS
the coherent three dimension superconducting state
the dependencies  $\xi _{ab} (T)$ and
$\xi_c (T)$ settles and the value $t_c$ doesn't depend on the temperature,
that is accorded with the London penetration
depth measurements [11].

  4. For quasi-2D HTS the temperature transition to
superconducting state $T_c$ with coherent dynamic of charge transport
along axis $\widehat{c}$ can be defined by
inequalities for layer systems (see [7,8], and the equation (12)
in review [12]):
\begin{equation}\label{1}
(\ln t_{c} (T_c))^{-1/3} \ge T_{c0}/ \varepsilon _F; \hspace{5truemm}
t_{á}('_c) \le T_{c0}/ \varepsilon _F,
\end{equation}

where $ \varepsilon _F$ the Fermi energy. As we can see from (1), the
value $T_c$ dependence from the values of three parameters:
$ t_c, T_{c0}, \varepsilon _F $ .  The
values of $ t_{c}$ and $T_{c0}$ can be found out at 
resistivity measurements
$\rho_c(T)$ and permit to define the relation between $T_c$ and
$ \varepsilon _F$. The comparison the results of the
resistivity measurements and the solution of the inequalities (1)
let us to qualify the pairing mechanism: or it is a mechanism described by
generalized BCS theory (such as discussed in Refs.[1-3]), or it is a
pairing mechanism that cannot be explained by the BCS theory
(such as spin analog of the superconducting proximity effect [13]).

Thus, in proposed variant superconductivity in
quasi-2D HTS assumes spin-fluctuational
pairing mechanism described by the generalized BCS theory
and occurs in two stages. At first at $T\sim T_{c0}$ strong 
superconducting fluctuations
are settled in copper-oxygen planes, 
which leads to essential temperature dependence of the
strength of interlayer coupling $t_{c}(T)$.
The transition of the sample to the coherent
superconducting state occurs at $T_c<T_{c0}$
at sufficiently small values of $t_{c}(T)$, which
fulfil to the inequalities (1).

References

1. A.Millis, P.Montoux, and D.Pines, Phys.Rev. B 42, 167 (1990)

2. G.G.Sergeeva, Yu.P.Stepanovskii, and A.V.Chechkin
   Low Temp. Phys. 24, 771 (1998)

3. Yu.A.Izyumov, Usp. Fiz. Nauk, 169, 265 (1999)

4. T.Watanabe and A.Matsuda, Phys.Rev. B 54, 6881 (1996)

5. T.Watanabe, T.Fujii, and A.Matsuda, Phys.Rev.Lett. 79, 2113 (1997)

6. G.G.Sergeeva, $M^{2}S-HTSC-YI$ contributed paper

7. E.I.Kats, Zh.Eksp. Teor. Fiz., 56, 1675 (1965)

8. L.N.Bulayevsky, Usp. Fiz. Nauk , 116, 449 (1975)

9. Y.Zha, Phys.Rev. B 53, 8253 (1996)

10. L.B.Ioffe, A.I.Larkin, A.A.Varlamov, and L.Yu.
    Phys.Rev. B 47, 8936 (1993)

11. S.Ushida, K.Tamasaki, and S.Tajima. Phys.Rev. B 53, 14558 (1996)

12. G.G.Sergeeva. Fiz. Nizk. Temp. 18, 797 (1992)

13. V.J.Emery, S.A.Kivelson, and O.Zachar. Phys.Rev. B 56, 6120 (1997)
\end{document}